\documentclass[12pt]{article}
\usepackage{float}
\usepackage[american]{babel}
\usepackage[utf8]{inputenc}
\usepackage{graphicx}
\usepackage{xcolor}
\usepackage[normalem]{ulem} 
\usepackage{makecell}

\usepackage[margin=1in]{geometry}
\usepackage{authblk}
\usepackage{amsmath}
\usepackage[
backend=biber,
style=nature,
sorting=none,
intitle=true,
maxnames=10
]{biblatex}
\usepackage[colorlinks=true,citecolor=blue,linkcolor=blue]{hyperref}

\newcommand{\ket}[1]{| #1 \rangle}

\newcommand{\ketbra}[2]{| #1 \rangle\langle #2 |}

\begin{document}

\title{Interactions of pre- and postselected quantum particles}

\author[1]{Gregory Reznik}

\affil[1]{\small School of Physics and Astronomy, Tel-Aviv University, Tel-Aviv 69978, Israel}
\author[2,3,4]{Jan Dziewior}
\affil[2]{Max-Planck-Institut f\"{u}r Quantenoptik, 85748 Garching, Germany}
\affil[3]{Fakult\"{a}t f\"{u}r Physik, Ludwig-Maximilians-Universit\"{a}t, 80797 M\"{u}nchen, Germany}
\affil[4]{\footnotesize Munich Center for Mathematical Philosophy, Ludwig-Maximilians-Universit\"{a}t, 80797 M\"{u}nchen, Germany}
\author[1,5]{Shrobona Bagchi}
\affil[5]{Quantum Universe Center, Korea Institute for Advanced Study, Seoul 02455, Korea}
\author[1,6]{Lev Vaidman}
\affil[6]{\small Institute for Quantum Studies, Chapman University, Orange, California 92866, USA}

\date{}
\maketitle

\begin{abstract}
An approach for analysis of effective interaction between pre- and postselected quantum particles is developed. It is argued that the cases of complete pre- and postselection of particles are more profound than the cases of partial pre- and postselection, since the former goes beyond modification of the average of interactions on an ensemble of experiments.
Recently discussed paradoxical phenomena such as the pigeonhole paradox and the modification of the interaction from repulsion to attraction are analyzed within the introduced formalism, and a few new surprising examples are presented.
\end{abstract}

\section{Introduction}
In this paper, we develop an approach for analyzing effective interactions between pre- and postselected quantum particles. 
The surprising  behavior of the particles in such cases has recently been reported \cite{Aharonov16,CCS}.
Here, we develop a more general analysis based on the combination of the locality of physical interactions and the following results: the description of the past of a pre- and postselected quantum particle according to the local trace it leaves \cite{past}, the universality property of the weak value of the projection operator on a particular location as a modifier of all effective interactions in this location \cite{PNAS}, a possibility of considering a weak value of a quantum particle with complete pre- and postselection as a property of a single particle \cite{beyond}, and a theorem connecting weak values of dichotomic variables with the certainty of results of strong measurements \cite{AV91}.
We show that weak values of local projections, products of local projections, and sums of products of local projections faithfully describe effective modifications of interactions between pre- and postselected particles.

In analyzing couplings of classical systems, it does not matter when the states of these systems were determined: before or after the interaction.
The full preselection or full postselection provides a complete description. This is not the case for quantum systems.
The coupling depends on both and is well described by the two-state vector formalism (TSVF) \cite{ABL,AV90}.
An extensive literature on this subject, typically considers a pre- and postselected quantum system which is coupled to another system which is preselected only.
Here we extend the analysis also to situations in which both systems (particles) are pre- and postselected.

A classical particle may or may not be in a particular location $A$.
In classical physics, we have no situations in which there is an objective uncertainty about the presence of the particle: we might not know it, but the particle is either present or not. 
There is a clear ontological meaning of the presence of a particle $i$ at a space-time point $(\vec{r}_A, t_0)$.
Every particle has a world line $\vec{r}_i(t)$ and the point $(\vec{r}_A, t_0)$ either belongs to this line or not. 

In addition to the ontological meaning, the presence of the particle in $(\vec{r}_A, t_0)$ has an operational meaning.
First, it corresponds to a counterfactual statement: If we ``look there'', i.e. perform a measurement sensitive to the presence of the particle at $A$, we will find the particle with certainty.
Furthermore, it is reasonable to assume that the particle will exhibit a non-vanishing (albeit possibly very tiny) local interaction with the environment, so apart from the counterfactual statement, there is also a factual statement: the states of objects in the vicinity of $\vec{r}_A$ are changed compared to an identical situation without the presence of the particle.
This change can be considered as a criterion for the presence of the particle.

If the wave function of a quantum particle at time $t_0$ is well localized in the vicinity of $A$, then the operational meaning of its presence is the same as for a classical particle, but for general wavefunctions, the question of the presence of a quantum particle is much more involved.
First, there is no consensus about the fundamental ontology of quantum theory.
Second, in the quantum case, there will be no certainty about what we will observe if we look for the particle at $A$ at time $t_0$, and third, the possible states of the local environment at $A$ do not have a simple dichotomic structure as in the classical case.

When considering a larger number of particles, in the classical case, the description of their separate presence answers all questions about their joint presence.
Since quantum theory is not separable, this is not true in the case of several quantum particles.
There are cases in which separate presence of particles does not ensure their joint presence, such as in the Hardy paradox \cite{hardy,product} in which, according to the operational meaning presented above, an electron is present in a particular place, a positron is present in the same place as well, but they are not present together.

Hardy's paradoxical situation occurs when we consider the properties of particles between two measurements.
Standard quantum mechanics avoids paradoxes by usually not discussing the question:
Where was the particle between two measurements? (See the recent discussion \cite{saldana2024,salcom,salrep}.)
The standard approach specifies the preparation and then provides a calculation of probabilities for the outcomes of subsequent measurements.
Between the measurements, at a time $t$ before the second measurement, when we imagine how things were when the second measurement has not yet been performed, the only available description of the particle would be its quantum state (the wave function).
In retrospect, however, at the time after the second measurement, when its result is known to us, we have access to more information than solely the quantum state fixed by the preparation.
It seems reasonable to use this additional information to describe the particle between the measurements. 

The first serious attempt to take into account the information from the postselection measurement was made by Wheeler \cite{Wheeler}, who suggested using this information to limit the description to the part of the wave function that reaches the postselection detector.
Vaidman \cite{past} proposed another way to take into account the postselection.
He suggested considering the traces, that is, the footprints the particle leaves on other systems at location $A$ due to the interaction at time $t$. 
All, even very weak, couplings to the environment are considered. According to Vaidman's proposal, the presence of a particle at a particular location at a particular moment of time is characterized
by these traces
compared to a reference trace defined by the footprints of a hypothetical particle with a well-localized wave function at $A$.
This characterization of the presence is very different from the concept of the presence of a classical particle and led to extensive theoretical and experimental research~\cite{LiCom,RepLiCom,morepast,Jordan,JordanCom,Danan,Bart,BartCom,Poto,PotoCom,Grif,GrifRep,Hash,HashCom,HashComRep,Dupr,DuprCom,DuprComRep,Eli,Disapp,Sokol2,ACWE,Saldan,Sali,SaliCom,Nik,NikRep,NikRR,China,ChinaCom,Sok,SokCom,SokComRep,Berge,BergeCom,BergeRep,Wiesn,Yuan1,Lemmel,Hance,HanceCom,Danner,Kim,vaidman2024,Bhati,PhotonsLyingAgain,Uranga,Dajka,Terris,Bernardo,YuanPhotons,YuanPhotonsCom,McQueen,Geppert,Waegell2023,Saeed,YuanHiding}.
Here we want to shed light on this research by conducting a more general analysis of Vaidman's proposal and connecting it to recent surprising quantum effects encountered in pre- and postselected systems.

In Section II we describe Vaidman's concept of the presence of a pre- and postselected particle \cite{past} that we adopt in our analysis.
In Section III we analyze several conceptually different cases of effective couplings of pre- and postselected particles to preselected systems clarifying the difference between two types of weak values which characterize the couplings: ``pure'' and ``mixed''.
Section IV utilizes the examples of Section III to analyze the effective coupling of one pre- and postselected particle to another pre- and postselected particle. 
In Section V we apply our formalism to several cases of peculiar behavior of pre- and postselected systems described in previous works.
Section VI provides a summary and concluding remarks.

\section{The Concept of Presence of a Pre- and Postselected Particle}

Vaidman's suggestion \cite{past} represents a conceptually radical change to the analysis of the question: Where was the particle?
Instead of proposing a picture of the particle with an ontological position and asking how this ontic entity moves in space, the presence of the particle is specified by the local trace it leaves on other systems.

In the current analysis, we consider quantum mechanics in which time is a parameter, so it is not problematic to consider an infinitesimally short time duration $\tau$.
There are known difficulties in dealing with very small volumes, i.e., {\it exact} locations, in quantum mechanics.
In typical considerations the presence of particles is evaluated with respect to a more coarsely grained spaces, such as paths through an interferometer.
Thus, we consider the presence of a particle at a particular time in a finite region of space.
To define the presence of the particle in this region, we consider its local couplings for a time $\tau \rightarrow 0$, where the only significant effects are those in the first order in $\tau$.

In general, the interaction leads to entanglement between the particle and the environment.
Moreover, the local environment even before the interaction can be in a mixed state.
However, via purification, it is possible to formulate all our results in terms of pure initial states.
This is done by considering a pure state of a larger system which includes the local environment and by considering degrees of freedom of the particle that become entangled with the local environment as part of the environment.

In the original theoretical analysis \cite{past}, a simplified model was considered in which the local interaction of the particle changes the state of the environment without changing the state of the particle. 
This is in contrast with experimental interferometric demonstration of the surprising consequences of Vaidman's approach~\cite{Danan} in which the effect of local interactions was observed on some degrees of freedom of the particle and not of the environment.
This represents a serious tension between the theoretical idea and experimental implementation: the theoretical proposal is to look at the effect on the environment, but what was observed is the effect on the particle.
Here, we will provide a more general analysis to clarify the controversy.

In interferometric scenarios such as~\cite{Danan} the particle enters a set of possible interferometric paths or ``arms'' it can traverse. 
The preselection corresponds to a particle entering a particular input port of the interferometer, so the preselected state is a superposition of localized wavepackets in different arms of the interferometer. The postselection corresponds to a particular output port that specifies only the ``arm'' degree of freedom of the particle. 
The general pre- and postselection states are 
\begin{align}
    \label{setup-pre}
    |\psi\rangle &= \sum_k \alpha_k|X_k\rangle, \\
    \label{setup-post}
    |\varphi\rangle &= \sum_k \beta_k|X_k\rangle,
\end{align}
where $|X_k\rangle$ signify arm $k$ of the interferometer at a particular time.

\begin{figure}[t]
\centering
\includegraphics[height=0.5\textwidth]{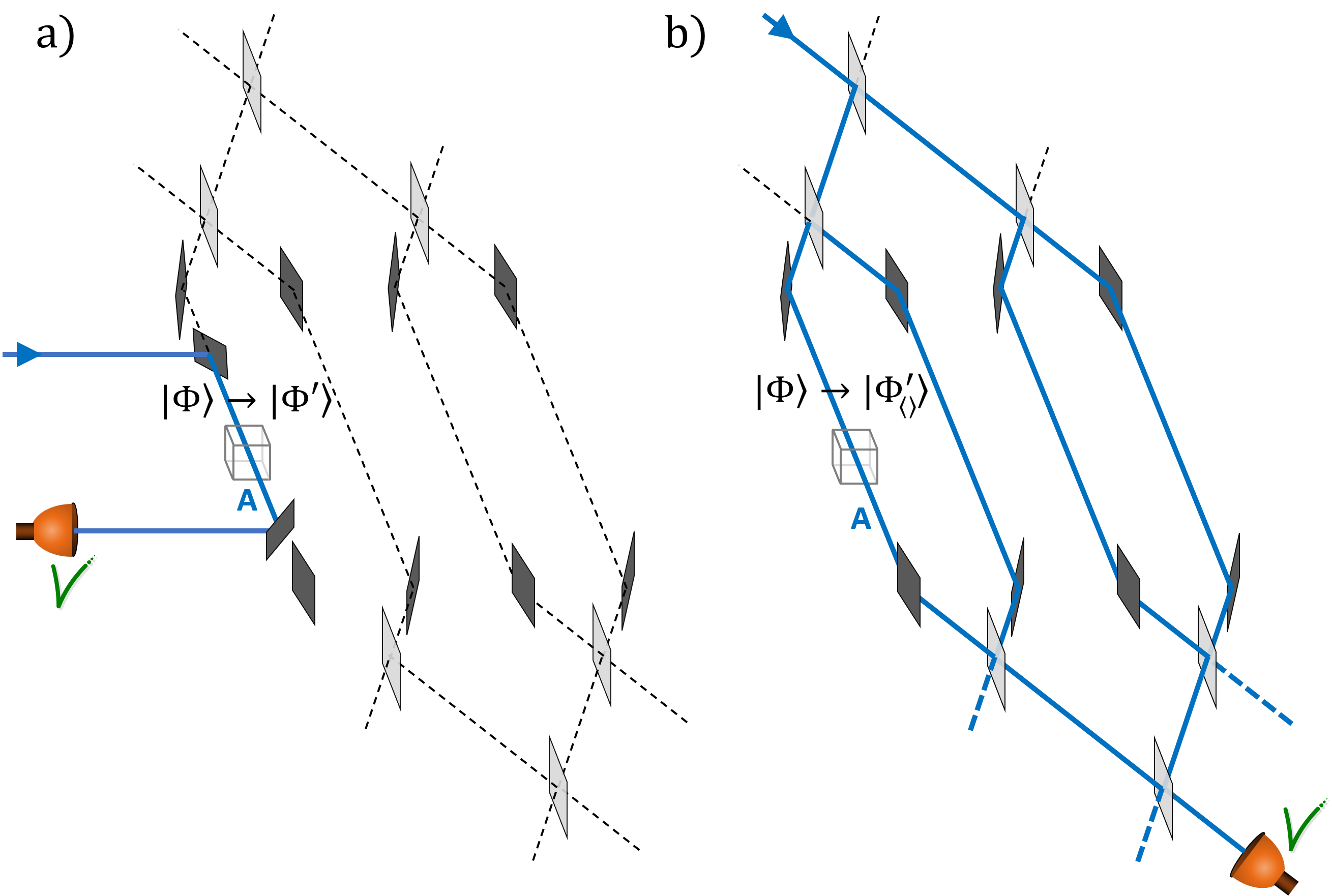}\\
\caption{ {\bf Scenarios of Particle Presence.}
Mirrors are shown in black, beam splitters in gray.
a) The particle passes only through arm $A$. This corresponds to the preselection of state $|A\rangle$ just before the interaction region and (or) the postselection of the same state $|A\rangle$ shortly afterward.
b) The particle is described by a superposition of interferometric paths/arms. The entry in a particular input port of the interferometer and the detection in the particular output port are equivalent to pre- and postselection states of the form (\ref{setup-pre}) and (\ref{setup-post}) at an intermediate time.
}
\label{setup}
\end{figure}

The presence in a particular location at a particular time is characterized by the trace left in this space-time region.
This trace includes the change of the local quantum state in this region and the change of the state of the non-path degrees of freedom of the particle during this short time.
To analyze the presence in a particular arm $X_k=A$ of the interferometer we consider one such coupling in arm $A$ for a short time $\tau$ (which will be taken to a vanishing limit) with coupling strength $g$: 
\begin{equation}\label{coup}
    H_{\rm int}=g{\mathbf{P}_A}O_A.
\end{equation}
We assume that the effect of interactions of the particle in other arms during this short time can be neglected.
 
To evaluate the presence of the particle for various pre- and postselected states based on the trace it left, it is necessary to characterize the reference trace left by a well-localized particle, Fig.~\ref{setup}a.
We signify the quantum state of the local environment and all degrees of freedom of the particle, except the arm degree of freedom, before the interaction, as $|\Phi\rangle$.
After the interaction (\ref{coup}), in the limit of small $\tau$, we obtain the reference state corresponding to the eigenvalue ${\mathbf{P}_A}=1$,  the standard presence of the particle:
\begin{equation}\label{reference}
    |\Phi'\rangle \equiv |\Phi'_{\rm ev1}\rangle = \mathcal{N}(|\Phi\rangle + \gamma \tau |\Phi_{\perp}\rangle),
\end{equation}
where $\langle \Phi |\Phi_{\perp}\rangle=0$, $\mathcal{N}$ denotes a normalization factor, and the parameter $\gamma$ depends on the coupling strength $g$ and the initial states of the systems that are coupled to the path degree of freedom.
This equation can be derived, for example, from a general formula \cite{AV90}, $O|\Phi\rangle = \bar{O}|\Phi\rangle+\Delta O |\Phi_{\perp}\rangle$  with $\bar{O} = \langle \Phi |O|\Phi\rangle$ and $\Delta O =\sqrt{\langle \Phi |O^2|\Phi\rangle- 
(\langle \Phi |O|\Phi\rangle)^2}$ .

The quantum state of the local environment (including the additional degrees of freedom of the particle) when the arm degree of freedom is pre- and postselected, see  Fig.~\ref{setup}b, is described by a superposition of the same states $\ket{\Phi}$ and $\ket{\Phi_\perp}$, but with a modified relative amplitude.
(We neglect the coupling to the non-path degrees of freedom of the particle outside of arm $A$.)
In the limit of small $\tau$, the modification is given by a simple formula
\begin{equation}\label{coupgeneralPrePost}
    |\Phi'_{\langle\phi||\psi\rangle}\rangle=\mathcal{N}(|\Phi\rangle +   ({\mathbf{P}_A})_w \gamma \tau |\Phi_{\perp}\rangle).
\end{equation}
where $({\mathbf{P}_A})_w$ is the weak value of the projection operator on arm $A$ for a particle preselected in state $|\psi\rangle$ and postselected in state $|\phi\rangle$
\begin{equation}\label{wv}
    ({\mathbf{P}_A})_w=\frac{\langle \varphi| {\mathbf{P}_A} |\psi\rangle}{\langle \varphi| \psi\rangle}.
\end{equation}

Another way to express this result is that the effective interaction Hamiltonian changing the state of the environment given the postselection is
\begin{equation}\label{effectiveinteraction}
    H^{\rm eff}_{\rm int}=g~  ({\mathbf{P}_A})_w O_A,
\end{equation}
so it is natural to take the weak value $ ({\mathbf{P}_A})_w$  as a definition of the presence of the particle in the arm $A$.
In particular, $({\mathbf{P}_A})_w=0$ corresponds to the absence of the particle, $({\mathbf{P}_A})_w=1$ corresponds to the standard presence of the particle, but $ ({\mathbf{P}_A})_w$ could be any complex number, corresponding to a richer concept of the presence of a quantum particle pre- and postselected in the spatial degree of freedom \cite{PNAS}.

Although the criterion is expressed using a weak value, a concept of the TSVF, the definition of the presence in spatial locations of a pre- and postselected particle is independent of the formalism.
It is based on the following operational definition:
\begin{quote}
 The particle was present in $A$ 
 if the trace left by the particle in $A$ is of the order of the trace left by a hypothetical particle with a localized wave packet in $A$.   
\end{quote}
Still, the TSVF is very helpful, but as was already mentioned in Sec.~VI of \cite{past}, the weak value $ ({\mathbf{P}_A})_w$ of the projection on $A$ provides a faithful estimate of this trace only when the postselection is restricted to the path degree of freedom.
When internal degrees of freedom are pre- and postselected,  as in a paradoxical situation of the quantum Cheshire Cat \cite{Cheshire}, we have to consider all local operators in $A$ acting on these degrees of freedom. Thus, the general definition in the framework of TSVF is the following.
\begin{quote}
 The particle was present in $A$ if the weak value of at least one local variable of the form $(O {\mathbf{P}_A})_w $ does not vanish.   
\end{quote}
Note that $ ({\mathbf{P}_A})_w $ is useful, too: it characterizes the modifications of local couplings to other (non postselected) degrees of freedom \cite{PNAS}.

We define the presence of the particle in $A$ according to the effect of local interactions there.
These interactions affect the local environment and also some degrees of freedom of the particle which are not postselected by the setup.
Experimentally, it has been much simpler to observe the trace left on these degrees of freedom because otherwise a conditional analysis has to be performed.
In our scheme, we assume that we know the trace left on these degrees of freedom of the particle passing through $A$ in the form of a well-localized packet.
We also make an assumption (which is not always fulfilled) that this trace is not changed (or changed in a known way) until the time the particle is observed.

Due to the highly challenging nature of experiments determining the footprints of particles on the local environments, the ``non-arm'' degrees of freedom of the particle have been used to assess the presence of the particle in the arms of interferometers.
In the experiments \cite{Danan,Zhou,Hasegawa,myphotonsneutrons}
the footprints of the interaction were written on the degrees of freedom of the particles that were kept undisturbed until and during the postselection measurement interaction.  (Note the surprising effects when an internal degree of freedom is disturbed differently on various paths of the interferometer \cite{prl2026}.)
The mathematical description of an internal degree of freedom that kept the footprint of a local interaction was identical to the description of the degree of freedom of the external environment interacting with the particle.
So, if the trace is left only on one external system, i.e., we ask the question of presence in a particular arm, then we do not expect any difference between analyzing the trace on the local environment or the trace left on the non-postselected degree of freedom of the particle.
However, when we use this degree of freedom to ask questions about presence in more than one location (and thus significant interaction with it takes place in more than one arm of the interferometer) the analysis should be different.

\section{Pure and mixed weak values}

Is a weak value a property of a single system or a property of an ensemble? 
A similar question is often posed about the quantum state of a system: Is it a description of a single system or of an ensemble?
Even for a pure quantum state, a widespread view is that it refers to an ensemble.
The reason for this is that if we are given a single copy of a pure quantum state, there is no way to determine what it is with certainty. However, we \textit{can} assign meaning to the pure quantum state of a single system.
If I {\it know} the quantum state, there is an operational meaning regarding the certain outcome of an experiment: measuring the projection on this state will succeed with certainty.
It is this certainty that we take to define what it means for a description or property to refer to a single quantum system.

If a quantum system is entangled with another system and we know its (mixed) state described by a density matrix, the situation is different.
There is no measurement to be performed on this system alone, that will succeed with certainty for a system described by this state.
Thus, according to our operational criterion, mixed weak values have to be regarded as a property of an ensemble. 

If we consider a value of an observable of a system in a pure state, then, following a similar approach, an eigenvalue is a property of a single system. This value will be obtained with certainty if the observable is measured. However, if the state is not an eigenstate of the variable, then, although it can be assigned a well-defined number: the expectation value, this number cannot be found in a single measurement and it does not characterize any measurement with a certain outcome. The only operational meaning of the expectation value is the average of the outcomes of the measurements performed on an ensemble. (If the quantum state is protected \cite{AVprotective} we can measure it on a single system \cite{PMitaly}, however, in protective measurements the expectation value is the property of the state together with its protection mechanism.)

The weak value was introduced as an outcome of the standard measurement procedure with weakened coupling \cite{AV90} which required conditional averaging on an ensemble, the meaning  adopted in a review \cite{WeakJordan}.
In cases of preselected quantum systems  without postselection or with partial postselection this meaning of the weak value as a statistical average is the only one which is available.
However, when we consider quantum systems with complete pre- and postselection, the weak value has a stronger non-statistical meaning.
It characterizes an experiment with an outcome which we obtain with certainty \cite{beyond}.
We name this type of weak value a ``pure weak value''.

\begin{figure}[t]
\centering
\includegraphics[width=0.5\textwidth]{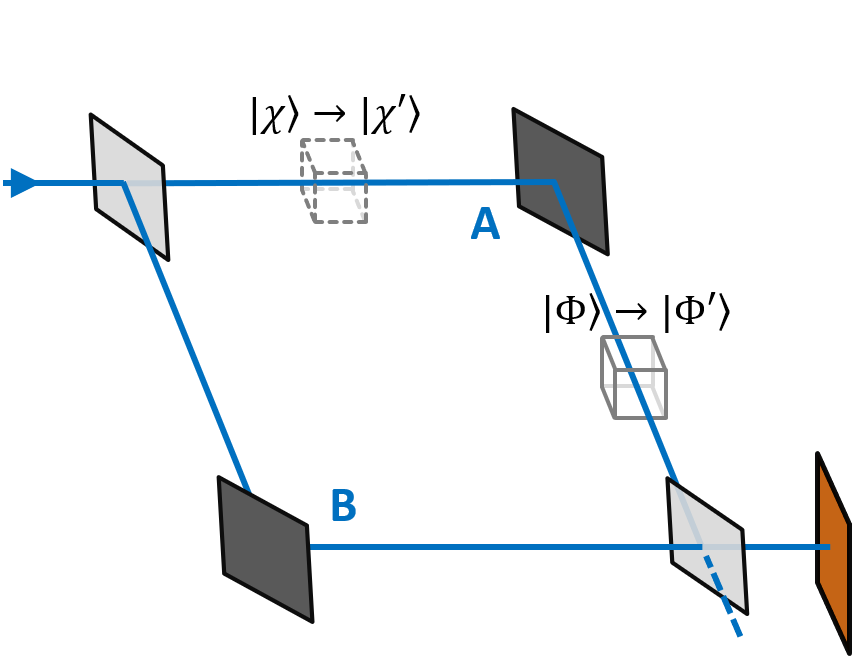}\\
\caption{ {\bf Mach-Zehnder Interferometer.}
The input and output ports are fixed.
Various cases are explored by varying the splitting-ratio and phase introduced by the first beam splitter, leading to different forward-evolving states of the form (\ref{sep-pre}).
The second beam splitter is balanced and has a fixed phase corresponding to the backward evolving state (\ref{sep-post}).
Apart from the short interaction constituting the weak trace, in some cases we also introduce decoherence through a controlled coupling to an external system with the initial state $|\chi\rangle$.}
\label{fig::MZI1}
\end{figure}

To demonstrate the difference between genuinely statistical average weak values and pure weak values we consider the paradigmatic case of an experiment with a single pre- and postselected particle passing through a Mach-Zehnder interferometer (MZI) with two arms $A$ and $B$, see Fig. \ref{fig::MZI1}.
We analyze the weak values of the projection operators on  the arms of the interferometer for several cases of pre- and postselection.
In each scenario, there are interactions with the mirrors and beam splitters of the MZI which determine the interferometric paths, a vanishingly short coupling of the weak measurement that probes the weak value, and, in some cases, decoherence due to additional controlled local couplings to an external system.

The imbalance of the input beam splitter is given by  the parameter $\alpha$ and the phase-shift $\phi$. The general expression for the preselected state is 
\begin{align}\label{sep-pre}
    |\psi\rangle = \cos \alpha |A\rangle + \sin \alpha ~e^{i\phi} |B\rangle.
\end{align}
In all cases the backward evolving state is postselected to be  
\begin{align}\label{sep-post}
    |\varphi\rangle = \frac{1}{\sqrt{2}}( |A\rangle + |B\rangle ).
\end{align}

The presence of the particle in arm $A$ is measured using the standard implementation of the ideal quantum measurement (\ref{coup})
where $O\equiv p$ is the momentum conjugate to the pointer variable $q$:
\begin{equation}\label{coup2}
H_{\rm int}=g{\mathbf{P}_A}~p.
\end{equation}
We choose the pointer of the measurement device in the initial Gaussian state $|\Phi\rangle$,
\begin{equation}\label{Gaus0}
\langle q | \Phi\rangle ={\mathcal{N}} e^{-\frac{q^2}{4\Delta^2}}
.\end{equation} 

The reference state describing the trace left by the particle passing through the arm for time $\tau$, is the state of the pointer of the measurement device shifted by $g\tau$ :
\begin{equation}\label{GausShift}
\langle q | \Phi'\rangle ={\mathcal{N}} e^{-\frac{(q-g\tau)^2}{4\Delta^2}},
\end{equation} 
which corresponds to $\gamma=\frac{g}{2\Delta}$ in (\ref{reference}).
This state is obtained when the particle is preselected in the path eigenstate $|A\rangle$
and an arbitrary postselected state.
Alternatively, we can have an arbitrary preselected state and postselection of the same eigenstate $|A\rangle$.

In all our analyses, we consider the selection of single input and output ports of the MZI,
so the pre- and postselection of the particle are complete. 
This leads to a pure weak value of projection, unless there is a decoherence coupling within the interferometer and we obtain a mixed weak value, see Fig. \ref{fig::MZI1}.
We will compare numerically equal pure and mixed weak values
for three  cases:
When the weak value equals one of the eigenvalues (1 or 0), when the weak value is real, but not equal to an eigenvalue, and finally, when the weak value is imaginary. 

We will start with pure weak values.
Without decoherence, the weak value of the projection for the particle to be in arm $A$ is
\begin{align}
    (\mathbf{P}_{A})_w=  \frac{1}{1+\tan \alpha e^{i\phi}}.
\end{align}

For a dichotomic variable, such as projection on the arm of the interferometer, a pure weak value equal to an eigenvalue is essentially equivalent to the case of the system being pre- or postselected in this eigenvalue.
Furthermore, for dichotomic variables, if a pure weak value equals an eigenvalue, then a strong measurement will yield this eigenvalue with certainty. 
For the initial state (\ref{sep-pre}) with $\alpha=0$, the weak value is   $ (\mathbf{P}_A)_w =1$. The coupling (\ref{coup2}) leads to the shifted pointer state (\ref{GausShift}) even for a finite coupling time $\tau$, and the expectation value of the quantum pointer is $\langle q \rangle = g\tau$.

In our first example, we consider a pure weak value which does not equal any eigenvalue, so a strong measurement cannot yield the same value.
The pointer will show one of the eigenvalues (or a statistical mixture of them according to some interpretational approaches).
Still, we will show below that in the limit of weak coupling, the measuring device ends up in a pure, merely shifted state, as if it were coupled to a hypothetical system in an eigenstate with an eigenvalue numerically equal to the weak value.

{\bf i)} We consider $ (\mathbf{P}_A)_w =2$ as an example of a weak value different from the allowed eigenvalues. It is obtained with an initial state (\ref{sep-pre}) having the parameters
    \begin{equation}
    \alpha_{\rm i}=\arctan\frac{1}{2},~~~~\phi_{\rm i}=\pi,
\end{equation}
and postselection of (\ref{sep-post}).
Indeed, the weak value in this case (no entanglement with external system and vanishingly small time $\tau$ of weak measurement, the decoherence effect of which can be neglected) is 
\begin{equation}\label{wvA}
(\mathbf{P}_{A}^{\rm i})_w=
\frac{\Big( \langle A| + \langle B|\Big) {\mathbf{P}_A} \Big(2|A\rangle - |B\rangle \Big)}
{\Big(\langle A| + \langle B|\Big) \Big(2|A\rangle - |B\rangle \Big)}=2. 
\end{equation}
The pointer state after the postselection is 
\begin{align}
|\Phi^{'}_{\rm i} \rangle =
 \mathcal{N} \Big(
  2  |\Phi' \rangle 
  -   |\Phi \rangle 
\Big)=
 {\mathcal{N}}(|\Phi\rangle +   2\gamma \tau |\Phi_{\perp}\rangle).
\end{align} 
The second equality is obtained in the limit of vanishingly small $\tau$ and it leads to the expectation value of the pointer
\begin{equation}
\langle q\rangle_{\rm i} = 
\langle \Phi^{'}_{\rm i} | q | \Phi^{'}_{\rm i}\rangle 
= 2g\tau.
\end{equation}

The final pointer state of the measuring device in a hypothetical experiment with a coupling to a system in an eigenstate corresponding to the eigenvalue 2, is:
\begin{equation}\label{ev2}
 \langle q| \Phi^{'}_{\rm ev2}\rangle = {\mathcal{N}} e^{-\frac{(q-2g\tau)^2}{4\Delta^2}}
.\end{equation} 
The final pointer state in our experiment (with eigenvalues 1 and 0) is very close to this state.
The Bures angle between these states is
\begin{align}\label{Dist4_5}
&D_A^{\rm i,ev2} =\arccos
|\langle \Phi^{'}_{\rm i}  |\Phi^{'}_{\rm ev2}   \rangle|
=  \frac{1}{2\sqrt{2}} \big(\frac{g\tau}{\Delta}\big)^2 +  O\big((\frac{g\tau}{\Delta})^3\big).
\end{align}

{\bf ii)} The second example of a pure weak value is $({\mathbf{P}_A})_w=i$.
It is instructive to see that the formalism can be applied in a similar way when weak values are imaginary, and it will be relevant in Section V as well.
For this weak value we preselect state (\ref{sep-pre}) with parameters
\begin{equation}\label{case_i_clean}
    \alpha _{\rm ii}=\arctan\sqrt{2},~~~~\phi_{\rm ii}=\frac{5\pi}{4},
\end{equation}
and employ the postselection (\ref{sep-post}).
The pointer state after the postselection is 
\begin{align}
|\Phi^{'}_{\rm ii} \rangle = {\mathcal{N}}
 \Big(
 |\Phi' \rangle -  (1+i) |\Phi \rangle 
\Big)=
 {\mathcal{N}}(-i|\Phi\rangle +   \gamma \tau |\Phi_{\perp}\rangle),
\end{align} 
the expectation value of the pointer variable vanishes,
\begin{equation}
\langle q\rangle_{\rm ii} = 
\langle \Phi^{'}_{\rm ii} | q | \Phi^{'}_{\rm ii}\rangle 
= 0,
\end{equation}
but the weak value  can be observed in the shift of the conjugate variable $\langle p\rangle_{\rm ii}=\frac{ \hbar g\tau}{2\Delta^2}$ , \cite{Jozsa}.

We want to argue that this pre- and postselected situation is very much the same as the case of an eigenvalue $i$.
However, there is no physical variable with such an eigenvalue. (In fact, the only way to obtain such an effective value is to use pre- and postselection.)  What we can do is to compare the state of the measuring device to the hypothetical state of a measuring device shifted by the ``eigenvalue'' $i$.
This reference state is
\begin{equation}
 \langle q| \Phi^{'}_{{\rm ev}\,i}\rangle = {\mathcal{N}} e^{-\frac{(q-i g\tau)^2}{4\Delta^2}}.
\end{equation} 
The Bures angle between the pointer states corresponding to a pure weak value $i$ and an ``eigenvalue'' $i$ is of second order in $\frac{g\tau}{\Delta}$ 
\begin{align}\label{Distii_i}
&D_A^{{\rm ii},{\rm ev}\,i} =\arccos
|\langle \Phi^{'}_{\rm ii} |\Phi^{'}_{{\rm ev}\,i}   \rangle|
= \frac{1}{4} \big(\frac{g\tau}{\Delta}\big)^2+ O\big((\frac{g\tau}{\Delta})^3\big).
\end{align}

{\bf iii)} Now we will turn to the cases of mixed weak values.
We will start with  $ (\mathbf{P}^{\rm iii}_A)_w =1$.
In spite of the fact that it is numerically equal to an eigenvalue, the case is not trivial. 
We preselect state (\ref{sep-pre}) with parameters
\begin{equation}\label{15}
    \alpha _{\rm iii}=\arctan\frac{1}{2},~~~~\phi_{\rm iii}=\pi,
\end{equation}
and postselect in state (\ref{sep-post}).


We add decoherence by coupling to an external system on path $A$.
We assume that without interaction with the particle, the final state of this system is $|\chi\rangle$.
The final state in the case the particle is well localized in $A$, is $|\chi'\rangle$.
Before the final beam splitter, the particle is entangled with this system, so postselection on a particular output port is a partial postselection of the composite system of the particle and the local external system. 
This postselection results in an entangled state of the measuring device and the local external system:
\begin{equation}\label{ex_1_pre+Phi}
{\mathcal{N}}\Big(
2  |\chi ' \rangle |\Phi'\rangle
  - |\chi \rangle |\Phi\rangle
\Big).
\end{equation} 
We can imagine (or actually perform) an additional measurement of the external system in some basis.
Then, for each outcome, we get a particular (different) pure weak value of the projection on $A$.
The weighted mixture of these outcomes provides the weak value of our experiment.
This is why we name this case a ``mixed weak value''.
 
An interesting choice is a measurement of the external system in a basis that includes $\{|\chi'\rangle,|\chi'_{\perp}\rangle\}$, where $\langle \chi'|\chi^\prime_\perp\rangle=0$, because it will create a particular mixture of weak values of projection on $A$, one of which is zero.
In fact, obtaining the state $|\chi'_{\perp}\rangle$ will lead to the final state $ |\Phi\rangle$ corresponding to
$(\mathbf{P}_{A}^{|\chi'_{\perp}\rangle})_w=0$. 
In order to achieve $ (\mathbf{P}^{\rm iii}_A)_w =1$, we need a relatively strong decoherence characterized by 
\begin{equation}\label{xhi|xhi}
\langle\chi|\chi'\rangle=\frac{1}{2}.
\end{equation}
For this value, when the hypothetical measurement of the environment ends up with the state  $|\chi'\rangle$, the state of the measuring device becomes
\begin{align}\label{wvforchi'}
{\mathcal{N}} \Big(
2  |\Phi'\rangle
  - \frac{1}{2} |\Phi\rangle
\Big)= {\mathcal{N}}(\frac{3}{2}|\Phi\rangle +   2\gamma \tau |\Phi_{\perp}\rangle),
\end{align} 
which corresponds to $(\mathbf{P}_{A}^{|\chi'\rangle})_w=\frac{4}{3}$.  The average shift of the pointer in a mixture is given by the sum of the shifts times their weights (probabilities) $\rm P$
\begin{equation}
\langle q \rangle_{\rm iii}=(\mathbf{P}_{A}^{|\chi'\rangle})_w g\tau {\rm P}({|\chi'\rangle}) + 0\cdot {\rm P}({|\chi'_{\perp}\rangle)} =g\tau ,
\end{equation}
which corresponds to 
$(\mathbf{P}_{A}^{~\rm iii})_w=1$.

The same result can be obtained without considering hypothetical measurements of the external system.
Tracing over the external system in the state (\ref{ex_1_pre+Phi}) yields a
 pointer  described by the mixture of shifted (\ref{GausShift}) and not shifted (\ref{Gaus0}) Gaussians
represented by the density matrix:

\begin{align}
    \label{ro'wv_1}
     \rho^\prime_{\rm iii} &= \mathcal{N} \Big( 4 \ketbra{\Phi^\prime}{\Phi^\prime} - \ketbra{\Phi^\prime}{\Phi} - \ketbra{\Phi}{\Phi^\prime} + \ketbra{\Phi}{\Phi} \Big) 
 ,\end{align}
which also yields $\langle q \rangle_{\rm iii}=g\tau$.

The weak value of the projection can also be obtained by considering the particle itself. Just before the postselection, the state of the particle and environment is 
\begin{equation}
\frac{1}{\sqrt{5}}\Big(
2 |A\rangle |\chi ' \rangle 
  -|B\rangle |\chi \rangle 
\Big).
\end{equation} 
Then the density matrix (in the basis $|A\rangle, |B\rangle$) of the particle is calculated by tracing out the environment degree of freedom and taking into account (\ref{xhi|xhi}):
\begin{equation}\label{intro-matx-mix}
\rho^{\rm pre}_{~\rm iii}= \frac{1}{5}
\begin{pmatrix}
4 & -1\\
-1 & 1
\end{pmatrix}
.\end{equation}
For mixed states, the weak value has been derived in \cite{beyond} (Eq.~(32) therein):
\begin{align} \label{eq::mixedWV1}
(\mathbf{P}_{A}^{~\rm iii})_w = \frac{\mathrm{Tr} \left( \rho^\text{post}_{~\rm iii} \mathbf{P}_A \rho^\text{pre}_{~\rm iii} \right)}{\mathrm{Tr} \left( \rho^\text{post}_{~\rm iii} \rho^\text{pre}_{~\rm iii} \right)}=1,
\end{align}
where (\ref{sep-post}) provides $\rho^\text{post}_{~\rm iii}=\frac{1}{ 2}\Big(|A\rangle + |B\rangle \Big)\Big(\langle A| + \langle B |\Big)$.

Although the expectation value of the pointer variable $\langle q \rangle_{\rm iii}=g\tau$
is the same as for the eigenvalue 1,
the pointer state is very different from 
(\ref{GausShift}).
The Bures angle between the mixed state of the pointer, described by ${\rho'}_{\rm iii}$, and  $|\Phi'  \rangle$ is 
\begin{align}\label{Bures_angle}
 D_A^{{\rm iii,ev1}} =\arccos
\sqrt{\langle \Phi'|{\rho'}_{\rm iii}|\Phi' \rangle}
=\frac{1}{\sqrt{12}}\frac{g\tau}{\Delta} + \mathcal{O}\Big(\big(\frac{g\tau}{\Delta}\big)^2\Big).
\end{align}
This distance is of  the same order as the distance between the initial and final states of the measuring device measuring the eigenvalue ${\mathbf{P}_A}=1$
\begin{align}
D^{\rm 0,ev1}_A =\arccos
|\langle \Phi|\Phi'  \rangle|= 
\frac{1}{2} \frac{g\tau}{\Delta} + \mathcal{O}\Big(\big(\frac{g\tau}{\Delta}\big)^2\Big).
\end{align}
This tells us that $ ({\mathbf{P}_A^{\rm iii}})_w=1  $ in case (iii) has a limited meaning.
It is very different from the particle actually being in the arm $A$.

The limitation of case (iii) looks even more significant with respect to arm $B$. As it has to be the case due to the additivity of weak values and presence of only single particle  inside the interferometer, $ (\mathbf{P}_B^{~\rm iii} )_w =0$.
This suggests that the particle is not present in $B$, but an infinitesimal measurement coupling in it leads to a change in the pointer comparable to the change in the pointer in the arm $A$.
Indeed, the coupling in arm $B$ leads to the pointer described by a mixed state 
\begin{align}
     \rho^{\prime B}_{\rm iii} &= \mathcal{N} \Big( 4 \ketbra{\Phi}{\Phi} - \ketbra{\Phi^\prime}{\Phi} - \ketbra{\Phi}{\Phi^\prime} + \ketbra{\Phi'}{\Phi'} \Big)
 .\end{align}
Although  $\langle q \rangle_{\rm iii}=0$,  the state of the pointer differs significantly from the state of the  undisturbed pointer and the Bures angle between them is
\begin{align}
D^{{{\rm 0,iii}^B}}_A = \arccos
\sqrt{\langle \Phi|{\rho'}_{\rm iii}^B|\Phi \rangle}= 
 \frac{1}{\sqrt{12}} \frac{g\tau}{\Delta} + O\big((\frac{g\tau}{\Delta})^2\big)
.
\end{align}

Another disadvantage of case (iii) is that an important theorem \cite{AV91} connecting weak and strong values does not hold.
For dichotomic variables, if a weak value equals an eigenvalue, a strong measurement will find this eigenvalue with certainty.
This is trivially true in the reference eigenvalue case $ {\mathbf{P}_A}=1$, but not in case (iii).
The probability of finding the particle in $A$ in a strong intermediate measurement is only $\frac{4}{5}$.

{\bf iv)} The next case is a mixed weak value $(\mathbf{P}_A^{\rm iv})_w=2$. We obtain it by 
preselecting the state (\ref{sep-pre}) with parameters
\begin{equation}
    \alpha _{\rm iv}=\arctan\frac{2}{3},~~~~\phi_{\rm iv}=\pi,
\end{equation}
and postselecting state (\ref{sep-post}).
The calculations, similar to case (iii) show that the mixed weak value $(\mathbf{P}_A^{\rm iv})_w=2$ (corresponding to $\langle q \rangle_{\rm iv}=2g\tau$) is obtained for a decoherence characterized by
\begin{equation}
\langle\chi|\chi'\rangle=\frac{17}{18}.
\end{equation}
Using the same method as in case (iii), we obtain the state of the pointer weakly coupled at $A$ after the postselection:
\begin{align}\label{Bures_density2}
     \rho^\prime_{\rm iv} &= \mathcal{N} \Big( 9 \ketbra{\Phi^\prime}{\Phi^\prime} - \tfrac{17}{3}\ketbra{\Phi^\prime}{\Phi} -\tfrac{17}{3} \ketbra{\Phi}{\Phi^\prime} + 4\ketbra{\Phi}{\Phi} \Big).
 \end{align}
The Bures angle between this state and the state of the pointer of the measurement device coupled to the eigenvalue 2, (\ref{ev2}) is:
\begin{align}
&D_A^{\rm iv,ev2} =\arccos
\sqrt{\langle  \Phi^{'}_{\rm ev2}|\tilde{\rho}^{'}_{\rm iv}|\ \Phi^{'}_{\rm ev2} \rangle}=
\sqrt{\frac{7}{20}} \frac{g\tau}{\Delta} + O\big((\frac{g\tau}{\Delta})^2\big)
.\end{align}
This is in contrast to the case of a numerically equal pure weak value (${\rm i}$).
The Bures angle between the final states of the pointer is of the next order of $g\tau/\Delta$ and therefore negligible, see (\ref{Dist4_5}).

{\bf v)} Finally, we consider a mixed weak value $({\mathbf{P}_A})_w=i$.
It can be obtained by preselection of
(\ref{sep-pre}) with 
\begin{equation}
    \alpha _{\rm v}=\arctan\sqrt{\frac{3}{2}},~~~~\phi_{\rm v}=\pi+\arctan\frac{1}{2},
\end{equation}
postselection of state (\ref{sep-post}) and decoherence characterized by 
\begin{equation}
\langle\chi|\chi'\rangle=\sqrt\frac{5}{6}.
\end{equation}

The pointer after the postselection is described by the density matrix
\begin{equation}
    \begin{aligned}
        \rho^\prime_\mathrm{v} &= \mathcal{N} \Big( 2 \ketbra{\Phi^\prime}{\Phi^\prime} + 3\ketbra{\Phi}{\Phi}  
        - (2+i) \ketbra{\Phi}{\Phi^\prime} - (2-i)\ketbra{\Phi^\prime}{\Phi} \Big),
    \end{aligned}
\end{equation}
which corresponds to the expectation value $\langle p\rangle_{\rm v}=\frac{\hbar g\tau}{2\Delta^2}$.

The expectation values of the pointer variable $q$ and the conjugate $p$ are the same for cases (ii) and (v), but the final states of the measuring devices are different.
The Bures angle between the pointer states of pure weak value $i$ and ``eigenvalue'' $i$ is of second order in $\frac{g\tau}{\Delta}$, see  
 (\ref{Distii_i}). In contrast, the Bures angle between the pointer states of the mixed weak value $i$ and the ``eigenvalue'' $i$ is of first order in $\frac{g\tau}{\Delta}$:
\begin{align}\label{Dist0_7}
&D_A^{{\rm v},{\rm ev}\,i} =\arccos
\sqrt{\langle \Phi^{'}_{{\rm ev}\,i} |\rho^\prime_\mathrm{v} |\Phi^{'}_{{\rm ev}\,i}   \rangle}
=  \frac{1}{2} \frac{g\tau}{\Delta} + O\big((\frac{g\tau}{\Delta})^2\big)
.\end{align}

\begin{table}
    \label{caseTable}
    \centering
    \bgroup
    \def\arraystretch{1.8}
    \begin{tabular}{ |c|c|c|c|c| } 
        \hline
        \makecell{case} & \makecell{type} & \makecell{weak value} & \makecell{coherence \\ $\langle \chi | \chi^\prime \rangle$} & \makecell{Bures angle \\ to ideal case} \\
        \hline
        \hline
        i & pure & 2 & 1 & $\frac{1}{2\sqrt{2}} \big(\frac{g \tau}{\Delta} \big)^2$ \\
        \hline
        ii & pure & $i$ & 1 & $\frac{1}{4} \big(\frac{g \tau}{\Delta} \big)^2$ \\
        \hline
        iii & mixed & 1 & $\frac{1}{2}$ & $\frac{1}{2} \frac{g \tau}{\Delta}$ \\
        \hline
        iv & mixed & 2 & $\frac{17}{18}$ & $\sqrt{\frac{7}{20}} \frac{g \tau}{\Delta}$ \\
        \hline
        v & mixed & $i$ & $\sqrt{\frac{5}{6}}$ & $\sqrt{\frac{5}{6}} \frac{g \tau}{\Delta}$ \\
        \hline
    \end{tabular}
    \egroup
    \caption{Overview of considered cases}
\end{table}

As summarized in Table~I, we described several cases: (i) and (ii) for pure weak values as well as (iii), (iv), and (v) for mixed weak values.
Pure weak values correspond to complete pre- and postselection.
In cases of pure weak values, the infinitesimal coupling to the measuring device leaves the latter in a pure state (after postselection), while for mixed weak values the measuring device ends up in a mixed state.

Pure weak values can always be verified with a vanishingly small probability of error.
Consider that we are given an unbound ensemble of identically pre- and postselected particles with a known weak value of some variable.
Then this pure weak value can be tested with arbitrary precision using measurements with definite outcomes such that the probability of even one failure can be made arbitrarily small.

If the weak value is not pure, it is mixed.
Such values can appear whenever the system is not completely pre- and postselected. Additional measurements before or after can change the weak value (and convert it to a pure one).
A mixed weak value cannot be tested with measurements having a vanishing probability of failure.
According to our definition stated in the beginning of this section, this implies that only pure weak values
can be thought of as
properties of single pre- and postselected quantum systems.

\section{Interaction between systems which are both pre- and postselected}

Until now, we considered coupling of the pre- and postselected particle to a measurement device which is preselected only.
However, when two pre- and (not fully) postselected particles are interacting with each other, observing their non-postselected degrees of freedom allows us to evaluate the effect of one particle on the other.

\begin{figure}[t]
\centering
\includegraphics[width=0.5\textwidth]{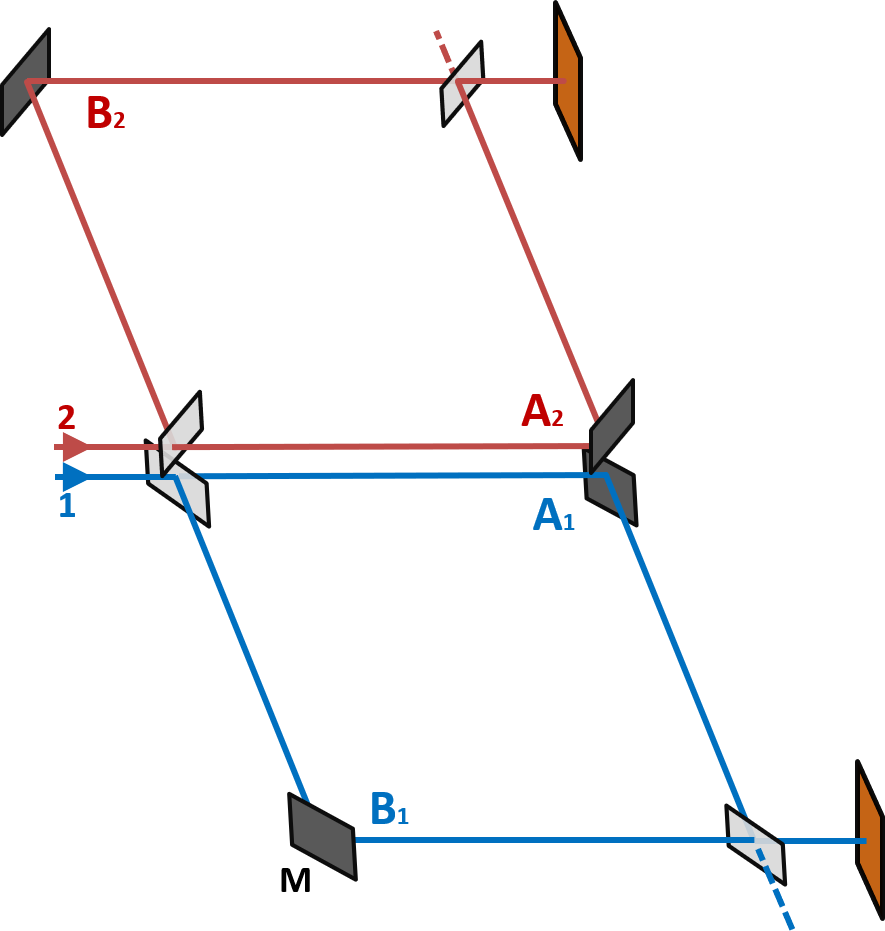}\\
\caption{ {\bf Two Mach-Zehnder Interferometers with interaction between particles in part of arm $A$.}
Two particles distinguishable by their transverse location pass simultaneously through interferometers close enough to have an observable interaction between them.
The transverse distance between the particles is much larger than the their transverse uncertainty.} 
\label{fig::MZI2AA}
\end{figure}

In the scenario we consider, each particle has its own interferometer.
We start by introducing an interaction (such as Coulomb force) only in one arm, so the setup can be described as in Fig.~\ref{fig::MZI2AA}.
The effect of the interaction can be measured by observing the shifts in transverse positions (the non-postselected degree of freedom) at the output port of the interferometers.
We assume that the uncertainty of the transverse position of the particles is much smaller than the separation between arms $A_1$ and $A_2$, so that we can neglect the entanglement of the transverse modes when the particles are only in $A$.
As before, for each of the particles, the initial state $| \Phi\rangle$ of the transverse position $q$ is the Gaussian (\ref{Gaus0}) and the reference shift is given by (\ref{GausShift}), which corresponds to the case when both particles are localized solely in the arm $A$.
The effective local interaction of the particles in arm $A$ when all arms of the interferometer are open can be measured by observing the modification of the transverse position shift at the output ports of the interferometers compared to the reference case.
The effective weak local interaction is modified by the weak value of the joint projection onto this location, $\mathbf{P}_{A1A2}={\mathbf{P}_{A1} \mathbf{P}_{A2}}$.

For particles which are pre- and postselected separately, the weak value of the product is equal to the product of the weak values.
This requires that there is no entanglement between the particles and also between the systems interacting with the particles.
Of course, an interaction between the particles creates entanglement, but here we consider a limit of weak coupling for which this entanglement can be neglected, such that the forward and backward evolving states can be considered product states between pre- and postselection and thus the product rule for weak values holds.
Thus, for separately prepared and detected particles, the modification of the interaction between them is fully described by their local presences:
\begin{equation}
\label{prod_rule}
 (\mathbf{P}_{A1} \mathbf{P}_{A2})_w=(\mathbf{P}_{A1})_w (\mathbf{P}_{A2})_w
.
\end{equation}

When we consider both particles pre- and postselected as in cases (i) and (iv) of Section 3, the effective coupling in $A$ will increase by the factor $2^2=4$.
For a reference case (with the particles both fully localized in $A$) in which the particles are shifted by $x$ (either repulsion or attraction), we thus expect to observe effective shifts of $4x$.
For particles pre- and postselected as in cases (ii) and (v), the modification factor is $i^2=-1$, so instead of a shift $x$ we will see the same shift in the opposite direction (turning repulsion to attraction and vice versa).

For some pairs of different pre- and postselections the spots on the detector observed for the two ensembles of particles can be very similar while the underlying pointer states are significantly different.
Straightforward calculations of the type performed in the previous section show that this is the case
for pairs of numerically equal pure and mixed weak values, see Table~I.

The state of each particle in a pure weak value case is identical, up to second order in $\big(\frac{g\tau}{\Delta}\big)$, to the state in the case of the numerically equal eigenvalue, while for mixed weak values the difference is of the first order.
In fact, 
the states differ only up to second order for case (i) and  eigenvalue 2 as well as case (ii) and ``eigenvalue'' $i$.
In contrast, for the mixed weak values (iii), (iv), and (v), the pointer states differ from the respective eigenvalue cases in first order.

The product rule for weak values (\ref{prod_rule}) holds only if there is no entanglement.
If the preselected or postselected (or both) states are entangled we might get surprising cases of the failure of the product rule \cite{nlexp,prodFailureExperimental}. 
For example, we can arrange a situation in which every particle is certainly inside $A$ but they are not present in $A$ together.
For this purpose, we use the following entangled pre- and postselected states of the two particles
\begin{align}
    |\psi\rangle = \frac{1}{\sqrt{3}} (|A\rangle_1|B\rangle_2  +  |B\rangle_1|A\rangle_2 - |B\rangle_1|B\rangle_2),
\end{align}
\begin{align}
    |\varphi\rangle = \frac{1}{\sqrt{3}} (|A\rangle_1|B\rangle_2 +  |B\rangle_1|A\rangle_2 + |B\rangle_1|B\rangle_2).
\end{align}
It is easy to see that for every particle $(\mathbf{P}_{A})_w=1$, however, it is obvious that $(\mathbf{P}_{A1} \mathbf{P}_{A2})_w=0$.
In this example, we have complete pre- and postselection and thus pure weak values.
Therefore, we can make a pair of surprising statements:
If we observe the presence of each particle in $A$, we will find each one with certainty.
However, we will not see an interaction between them, as if they were never together in $A$. 

Not less paradoxical is the situation in which if we look for each particle in $A$ separately then we are certain not to find any of them, however, we will see that they show shifted pictures on the final detector as if they had interacted in arm $A$.
In this case we are also certain to find them if we are looking for the two together.
This can be achieved for example by the following pre- and postselection
\begin{align}
    |\psi\rangle = \frac{1}{\sqrt{7}} (|A\rangle_1 |A\rangle_2 - |A\rangle_1 |B\rangle_2 -  |B\rangle_1 |A\rangle_2 +  2|B\rangle_1 |B\rangle_2 ),
\end{align}
\begin{align}
    |\varphi\rangle = \frac{1}{2} (|A\rangle_1 |A\rangle_2 +  |A\rangle_1 |B\rangle_2 + |B\rangle_1 |A\rangle_2+ |B\rangle_1 |B\rangle_2)
.\end{align}

\begin{figure}[t]
\centering
\includegraphics[width=0.5\textwidth]{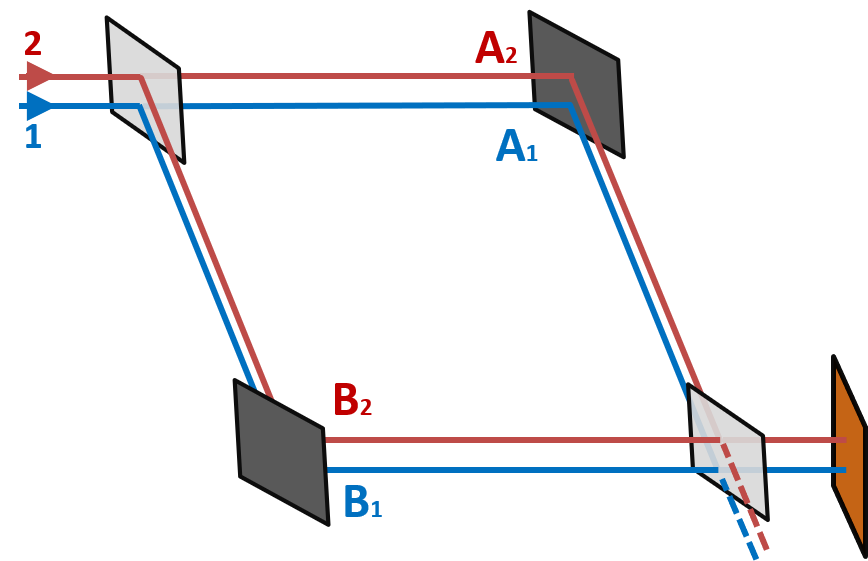}\\
\caption{ {\bf Single Mach-Zehnder Interferometer with interaction between particles in both arms.}
Two electrons distinguishable by their transverse location pass simultaneously through a Mach-Zehnder interferometer close enough to have observable interaction between them.
They interact in arm A and in arm B.} 
\label{fig::MZI2with1}
\end{figure}

So far we have considered coupling only in one arm of the interferometer.
It is of interest to consider also the case in which the coupling between particles is the same in all arms.
This corresponds to particles moving in the same interferometer with a small displacement see Fig.~\ref{fig::MZI2with1}. 
In this case the effective coupling in the output port will be described by the weak value of joint projection of the two particles in any of the two arms:
\begin{equation}\label{1with2_}
\mathbf{P}_{\rm 1with2}={\mathbf{P}_{A1} \mathbf{P}_{A2}}+{\mathbf{P}_{B1}\mathbf{P}_{B2}}.
\end{equation}
We will consider examples of such situations in the next section.

\section{Three-box paradox, pigeonhole paradox, electrons attracting each other }

In this section, we would like to connect our results with previous works.
First, consider the three-box paradox \cite{AV91}.
We prepare a particles in the state
\begin{equation}\label{three-box_pre}
  |\psi\rangle = \frac{1}{\sqrt{3}}( |A\rangle + |B\rangle - |C\rangle),  
\end{equation}
and postselect the state
\begin{equation}\label{three-box_post}
  |\varphi\rangle = \frac{1}{\sqrt{3}}( |A\rangle + |B\rangle + |C\rangle).  
\end{equation}

The weak values of projections on different boxes are
$(\mathbf{P}_{A})_w=(\mathbf{P}_{B})_w=1$ and $(\mathbf{P}_{ C})_w=-1$.
These are pure weak values, so if we are looking in box A we will find it with certainty, and if we are looking in box B instead, we will find it there with certainty too.
If we arrange a weak interaction in $A$, the beam in the output port will be shifted as if there was just one arm with the same coupling.
The observed effect does not change if such a coupling is present in all arms $A$, $B$, and $C$.
However, if the coupling is present, e.g., only in arms $A$ and $B$ but not in $C$, we will observe the sum of shifts which corresponds to an amplification by the factor of 2.

If we have two particles as in three-box paradox, with preselection (\ref{three-box_pre}) and postselection  (\ref{three-box_post}), and we arrange a weak coupling between them in all three boxes, the effective coupling will be three times bigger
\begin{equation}
(\mathbf{P}_{\rm 1with2})_w=
(\mathbf{P}_{A})_w^2 +
(\mathbf{P}_{B})_w^2+
(\mathbf{P}_{C})_w^2=3
.\end{equation}
This is directly explained when we understand that
we will not only find the two particles with certainty if we look for them together in $A$ or together in $B$, but also together in $C$, because
\begin{equation}
(\mathbf{P}_{1C} \mathbf{P}_{2C})_w=
(\mathbf{P}_{C})_w^2=(-1)^2=1.\end{equation}

Another peculiar scenario is the example of four particles pre- and postselected as in case (ii) of Section III.
The probability to find each particle when measuring its presence in $A$ is given by 
the Aharonov-Bergmann-Lebowiz rule \cite{ABL}
\begin{align}
\mathrm{P} (A)=
\frac{|\langle\varphi|\mathbf{P}_{A}|\psi\rangle|^2}
{|\langle\varphi|\mathbf{P}_{A}|\psi\rangle|^2+|\langle\varphi|\mathbf{P}_{B}|\psi\rangle|^2}
=\frac{|i|^2}{|i|^2+|1-i|^2}=\frac{1}{3}
.\end{align}
However, the probability of finding them together in $A$ is 1!
Indeed:
\begin{equation}
(\mathbf{P}_{A1}\mathbf{P}_{A2}\mathbf{P}_{A3}\mathbf{P}_{A4})_w=
(\mathbf{P}_{A})_w^4=(i)^4=1
.\end{equation}
Note that an experimental verification of this claim (as well as the claim of probability 1 to find pair of particles in box $C$ in the three-box paradox setup) is extremely difficult.
One cannot simply measure the number of particles in the box.
We have to devise a non-demolition experiment testing the projection on the state of four particles being present in the box. The procedure should not distinguish between three, two, one, or zero particles in the box. 

The next example is the quantum pigeonhole paradox first introduced in \cite{Aharonov16}.
We consider many particles in two boxes preselected in state (\ref{sep-pre}) with
\begin{equation}\label{pigeon}
    \alpha =\frac{\pi}{4},~~~~\phi=\frac{\pi}{2},
\end{equation}
and postselected in (\ref{sep-post}).
(We ``flipped'' pre- and postselection of  \cite{Aharonov16} to make it similar to other examples.)  Here, there is a surprising property that  particles from every particular pair  $\{\rm i,j\}$ are not  together. 
Indeed, for every particle
\begin{align}
({\mathbf{P}_{A}})_w=\frac{1}{1+i},
~~~~~~~~~
({\mathbf{P}_{B}})_w=\frac{i}{1+i}.
\end{align}
Thus, since all particles are pre- and postselected separately,
\begin{align} \label{Prij}
(\mathbf{P}_{\rm i\,with\,j})_w =
({\mathbf{P}_{A\rm i}\mathbf{P}_{A\rm j}})_w+ ({\mathbf{P}_{B\rm i}\mathbf{P}_{B\rm j}})_w=
({\mathbf{P}_{A\rm i})_w(\mathbf{P}_{A\rm j}})_w+ ({\mathbf{P}_{B\rm i})_w(\mathbf{P}_{B\rm j}})_w
=0.
\end{align}
The projection (\ref{Prij}) is a dichotomic variable, the result of strong measurement should also yield 0.
Many particles are placed into two boxes, but no two particles can be  found together in a measurement which only tests if they are together or not, without specifying the box.
Since physical interactions are bipartite, it means that we effectively remove the interaction between the particles in the boxes.
If we put several electrons in close parallel trajectories as in Fig.~\ref{fig::MZI2with1}, we do not expect any displacements on the position detectors in the ports corresponding to postselections of states $\prod_i |\varphi_i\rangle$ where all states $|\varphi_i\rangle$ are given by (\ref{sep-post}).

Our last example is inspired by Cenni,  Corr{\^e}a, and Saldanha (CCS) \cite{CCS}. They considered pre- and postselected electrons passing through an interferometer as in Fig.~\ref{fig::MZI2with1}, which demonstrated mutual attraction instead of repulsion.
We can achieve this task by preselecting two electrons in the state
\begin{equation}\label{CCS_pre}
|\psi\rangle=\frac{1}{2}
\Big( e^{i\phi}|A\rangle_1 + i|B\rangle_1 \Big)
\Big( ie^{i\phi}|A\rangle_2 + |B\rangle_2 \Big)
,\end{equation}
with $\phi = \frac{4\pi}{3}$  and postselecting the state 
 \begin{equation}\label{CCS_post}
 |\varphi\rangle=\frac{1}{2}
 \Big(-i|A\rangle_1 + |B\rangle_1 \Big)
 \Big(|A\rangle_2 -i |B\rangle_2 \Big).
\end{equation}
The repulsion of electrons can be modeled as a mutual measurement interaction (\ref{coup2}), so each electron can be considered as a measuring device for the other.  Given that the change of the forward and backward evolving states between pre- and postselection can be neglected, i.e., $g\tau \ll \Delta$, we obtain
\begin{equation}\label{weakCalc2}
(\mathbf{P}_{\rm 1with2})_w=\frac{\langle \varphi| {\mathbf{P}_{A1} \mathbf{P}_{A2}}+{\mathbf{P}_{B1}\mathbf{P}_{B2}} |\psi\rangle}{\langle \varphi| \psi\rangle}
=-1.
\end{equation}
Since this is a case of a pure weak value, in the limit of weak coupling the particles behave as if we would have an electron and a positron instead of two electrons.

However, CCS applied coupling to a measurement device with a finite coupling strength characterized by $g\tau =0.3\sqrt{2}\Delta$ ($\delta=0.3W$ in their notation) which causes significant entanglement between paths and transverse modes of the electrons. 
Thus, the scalar products between the states of the transverse modes of the electrons with and without the interaction given by (\ref{GausShift}) and (\ref{Gaus0}), yield the value:
\begin{equation}\label{CCS_overlap}
\langle \Phi_1' | \Phi_1\rangle=\langle \Phi_2' | \Phi_2\rangle=
e^{-\frac{(0.3)^2}{4}}
.\end{equation} 
The preselected  state of CCS had the form (\ref{CCS_pre}) but with a different phase, $\phi=\frac{3\pi}{4}$. 
Then, just before the postselection, the state of  the two electrons is
\begin{align} 
    \frac{1}{2} \Big(|A\rangle_1 |A\rangle_2 +
 i|B\rangle_1 |B\rangle_2 \Big)
 |\Phi _1 '\rangle |\Phi _2 '\rangle  +
\frac{e^{i\frac{3\pi}{4}}} {2} \Big( |A\rangle_1 |B\rangle_2 -
|B\rangle_1 |A\rangle_2 \Big)
|\Phi _1 \rangle  |\Phi _2 \rangle 
.\end{align} 
The density matrix of the system in the basis $\{|A_1 A_2\rangle , |A_1 B_2\rangle , |B_1 A_2\rangle, |B_1 B_2\rangle \}$  is calculated by tracing out the transverse degrees of freedom and taking into account (\ref{CCS_overlap}):
\begin{align}\label{rhonew}
\rho=\frac{1}{4}
\begin{pmatrix}
{1} & {e^{\frac{5\pi i-0.18}{4}}} & {e^{\frac{\pi i -0.18}{4}}} & {-i}\\
{e^{\frac{3\pi i-0.18}{4}}}  & {1} & -{1} &   {e^{\frac{\pi i-0.18}{4}}}\\
{e^{\frac{-\pi i-0.18}{4}}}  & -{1} & {1} &   {e^{\frac{5\pi i-0.18}{4}}}\\
{i} & {e^{\frac{-\pi i-0.18}{4}}} & {e^{\frac{3\pi i-0.18}{4}}} &  {1}
\end{pmatrix}
\end{align}
The postselection was on state $|\varphi\rangle$ given by (\ref{CCS_post}).
The weak value of the projection on being together in any of the arms (\ref{1with2_}), is constant during the time between pre- and postselection and thus can be calculated just before postselection:
\begin{align}\label{eq::wvModified}
(\mathbf{P}_{\rm 1with2})_w 
=\frac{\mathrm{Tr} \left( |\varphi_1\rangle  |\varphi_2 \rangle \langle \varphi_2|\langle\varphi_1| \mathbf{P}_{\rm 1with2} ~\rho \right)}{\mathrm{Tr} \left( |\varphi_1\rangle  |\varphi_2 \rangle \langle \varphi_2|\langle\varphi_1|  \rho \right)}
=-1.19.
\end{align}
This appears like a change of repulsion to attraction with a slight amplification.
By choosing $\phi=2.21$ instead of $\phi=\frac{3\pi}{4}$ we can obtain $(\mathbf{P}_{\rm 1with2})_w= -1$.
However, this does not exactly represent a switching of the interaction from repulsion to attraction.
In the CCS example the weak value is mixed, so it rather corresponds to a switching from repulsion to a mixture of different interactions that on average corresponds to attraction of the same strength as the original repulsion.
Only the pure weak value $(\mathbf{P}_{\rm 1with2})_w$ achieves an effective switching of the interaction from repulsion to attraction (in the limit of very short interaction).

\section{Conclusions}

We have revisited the concepts of the TSVF which is helpful for the analysis of pre- and postselected quantum systems.  
To avoid any confusion, we stress that the TSVF is fully compatible with the standard approach to quantum mechanics.
However, it simplifies calculations of effects of pre- and postselected quantum systems and provides intuition about what can be achieved using pre- and postselection.
It achieves this by describing the pre- and postselected system itself, while the standard calculation requires analyzing the system and its entanglement with other systems it interacted with between pre- and postselection. 

The main concept of the TSVF, the weak value, was introduced and is still frequently used as the outcome of a weakened measurement obtained by averaging on the pre- and postselected ensemble.
We extended this 
to the observation of the coupling between two systems which are \textit{both} pre- and postselected.

Furthermore, we argue that a more profound concept of a weak value is given when pre- and postselection is complete, which we named ``pure weak values'' (in contrast to mixed weak values emerging in cases where the pre- and postselection is not complete).
In the limit of very short coupling, the system interacts with any other system as if the weak value were an eigenvalue.
The deviation is only of the second order in the small parameter of the duration of the interaction, so the cases of numerically equal pure weak values correspond to essentially the same interaction even if they correspond to different pre- and postselections.
For mixed weak values the effect of the interactions of numerically equal weak values differ, in general, in the first order of the duration of the interaction, which is the same as the effect of the interaction itself, so we cannot claim that a mixed weak value fully specifies the interaction of a pre- and postselected system. 
Also, the theorem about dichotomic variables that predicts a probability of 1 for the outcome of a strong measurement when the weak value is equal to one of the eigenvalues holds only for pure weak values.
Since a projection is a dichotomic variable, the theorem allows
us to analyze several surprising effects regarding the presence and local interaction of pre- and postselected particles. 

Single particle weak values also provide an intuitive picture for the interaction between separately pre- and postselected particles.
More surprising effects (which are, however, much more challenging to demonstrate experimentally) occur when entanglement is present.
Then the weak value of the joint projection of two particles is not the product of the weak values of projections of each particle separately, and some counterintuitive situations occur.
Finally, our formalism sheds new light on recently discussed ``paradoxes'' related to pre- and postselected quantum systems.
We have shown how these paradoxes can be explained through our approach based on weak values of local projections, weak values of products of local projections, and weak values of the sums of these products.
The approach also helped us find some new paradoxical situations.
We hope that our analysis can lead to a deeper understanding of our quantum world and useful applications in quantum technology.

This work has been supported in part by the Israel Science Foundation Grant  No.~2689/23 and by PBC PostDoctoral Fellowship. We also acknowledge funding by the German Federal Ministry of Research, Technology and Space (Bundesministerium für Forschung, Technik und Raumfahrt, BMFTR) within the project QuKuk (Contract No. 16KIS1621), by the German Research Foundation (Deutsche Forschungsgemeinschaft, DFG) under Germany’s Excellence Strategy—EXC-2111—390814868 and by the Bavarian Ministry for Science and the Arts under the project EQAP (CR 20211118). SB is supported by KIAS individual
grant number QP100101 at Korea Institute for Advanced Study, Seoul, South Korea.

\printbibliography

@article{nlexp,
  title = {Measurements of Nonlocal Variables and Demonstration of the Failure of the Product Rule for a Pre- and Postselected Pair of Photons},
  author = {Xu, X. and Pan, W. and Wang, Q. and Dziewior, J. and Knips, L. and Kedem, Y. and Sun, K. and Xu, J. and Han, Y. and Li, C. and Guo, G. and Vaidman, L.},
  journal = {Phys. Rev. Lett.},
  volume = {122},
  pages = {100405},
  numpages = {7},
  year = {2019},
  publisher = {American Physical Society},
}

@article{past,
  title={Past of a quantum particle},
  author = {Vaidman, L.},
  journal={Phys. Rev. A},
  volume={87},
  pages={052104},
  year={2013},
  publisher={American Physical Society}
}

@article{CCS,
  title={Effective electrostatic attraction between electrons due to quantum interference},
  author = {Cenni, M. and Corr{\^e}a, R. and Saldanha, P.},
  journal={Phys. Rev. A},
  volume={100},
  pages={022101},
  year={2019},
  publisher={APS}
}

@article{Cheshire,
  title={Quantum {C}heshire cats},
  author = {Aharonov, Y. and Popescu, S. and Rohrlich, D. and Skrzypczyk, P.},
  journal={New J. Phys.},
  volume={15},
  pages={113015},
  year={2013},
  publisher={IOP Publishing}
}

@incollection{Wheeler,
  title={The “past” and the “delayed-choice” double-slit experiment},
  author = {Wheeler, J.},
  booktitle={Mathematical foundations of quantum theory},
  pages={9--48},
  year={1978},
  publisher={Elsevier},
}

@article{Danan,
  title={Asking photons where they have been},
  author = {Danan, A. and Farfurnik, D. and Bar-Ad, S. and Vaidman, L.},
  journal={Phys. Rev. Lett.},
  volume={111},
  pages={240402},
  year={2013},
  publisher={American Physical Society},
}

@article{Aharonov16,
  title={Quantum violation of the pigeonhole principle and the nature of quantum correlations},
  author = {Aharonov, Y. and Colombo, F. and Popescu, S. and Sabadini, I. and Struppa, D. and Tollaksen, J.},
  journal={Proc. Natl. Acad. Sci. USA},
  volume={113},
  pages={532},
  year={2016},
  publisher={National Academy of Sciences}
}

@article{PNAS,
  title={Universality of local weak interactions and its application for interferometric alignment},
  author = {Dziewior, J. and Knips, L. and Farfurnik, D. and Senkalla, K. and Benshalom, N. and Efroni, J. and Meinecke, J. and Bar-Ad, S. and Weinfurter, H. and Vaidman, L.},
  journal={Proc. Natl. Acad. Sci. USA},
  volume={116},
  pages={2881},
  year={2019},
  publisher={National Acad Sciences}
}

@Article{beyond,
  author = {Vaidman, L. and Ben-Israel, A. and Dziewior, J. and Knips, L. and Wei\ss{}l, M. and Meinecke, J. and Schwemmer, C. and Ber, R. and Weinfurter, H.},
  title     = {Weak value beyond conditional expectation value of the pointer readings},
  journal   = {Phys. Rev. A},
  year      = {2017},
  volume    = {96},
  pages     = {032114},
  issue     = {3},
  numpages  = {13},
  publisher = {American Physical Society},
}

@article{AV91,
  title={Complete description of a quantum system at a given time},
  author = {Aharonov, Y. and Vaidman, L.},
  journal={J. Phys. A: Math. Gen.},
  volume={24},
  pages={2315},
  year={1991},
  publisher={IOP Publishing}
}

@article{saldana2024,
  title={Association between quantum paradoxes based on weak values and a realistic interpretation of quantum measurements},
  author = {Aredes, A. M. and Saldanha, P. L.},
  journal={Phys. Rev. A},
  volume={109},
  pages={022238},
  year={2024},
  publisher={APS}
}

@article{product,
  title = {{L}orentz-invariant ``elements of reality'' and the joint measurability of commuting observables},
  author = {Vaidman, L.},
  journal = {Phys. Rev. Lett.},
  volume = {70},
  issue = {22},
  pages = {3369},
  numpages = {0},
  year = {1993},
  publisher = {American Physical Society},
}

@article{hardy,
  title={Quantum mechanics, local realistic theories, and {L}orentz-invariant realistic theories},
  author = {Hardy, L.},
  journal={Phys. Rev. Lett.},
  volume={68},
  pages={2981},
  year={1992},
  publisher={APS}
}

@article{Zhou,
  title={Experimental observation of anomalous trajectories of single photons},
  author = {Zhou, Z. and Liu, X. and Kedem, Y. and Cui, J. and Li, Z. and Hua, Y. and Li, C. and Guo, G.},
  journal={Phys. Rev. A},
  volume={95},
  pages={042121},
  year={2017},
  publisher={APS}
}

@article{AVprotective,
  title={Measurement of the {S}chr{\"o}dinger wave of a single particle},
  author = {Aharonov, Y. and Vaidman, L.},
  journal={Phys. Lett. A},
  volume={178},
  pages={38},
  year={1993},
  publisher={Elsevier}
}

@article{PMitaly,
  title={Determining the quantum expectation value by measuring a single photon},
  author = {Piacentini, F. and Avella, A. and Rebufello, E. and Lussana, R. and Villa, F. and Tosi, A. and Gramegna, M. and Brida, G. and Cohen, E. and Vaidman, L. and Degiovanni, I. and Genovese, M.},
  journal={Nat. Phys.},
  volume={13},
  pages={1191},
  year={2017},
  publisher={Nature Publishing Group UK London}
}

@article{WeakJordan,
  title={Colloquium: Understanding quantum weak values: Basics and applications},
  author = {Dressel, J. and Malik, M. and Miatto, F. M. and Jordan, A. N. and Boyd, R. W.},
  journal={Rev. Mod. Phys.},
  volume={86},
  pages={307},
  year={2014},
  publisher={APS}
}

@article{AV90,
  title={Properties of a quantum system during the time interval between two measurements},
  author = {Aharonov, Y. and Vaidman, L.},
  journal={Phys. Rev. A},
  volume={41},
  pages={11},
  year={1990},
  publisher={APS}
}

@article{Hasegawa,
  title={Multifold paths of neutrons in the three-beam interferometer detected by a tiny energy kick},
  author = {Geppert-Kleinrath, H. and Denkmayr, T. and Sponar, S. and Lemmel, H. and Jenke, T. and Hasegawa, Y.},
  journal={Phys. Rev. A},
  volume={97},
  pages={052111},
  year={2018},
  publisher={APS}
}

@article{myphotonsneutrons,
  title={Neutrons and photons inside a nested {M}ach-{Z}ehnder interferometer},
  author = {Vaidman, L.},
  journal={Phys. Rev. A},
  volume={101},
  pages={052119},
  year={2020},
  publisher={APS}
}

@article{Jozsa,
  title = {Complex weak values in quantum measurement},
  author = {Jozsa, R.},
  journal = {Phys. Rev. A},
  volume = {76},
  pages = {044103},
  numpages = {3},
  year = {2007},
  publisher = {American Physical Society},
}

@article{LiCom,
  title={Comment on “{P}ast of a quantum particle”},
  author = {Li, Z. and Al-Amri, M. and Zubairy, M. S.},
  journal={Phys. Rev. A: At. Mol. Opt. Phys.},
  volume={88},
  pages={046102},
  year={2013},
  publisher={APS}
}

@article{RepLiCom,
  title={Reply to “{C}omment on ‘{P}ast of a quantum particle’”},
  author = {Vaidman, L.},
  journal={Phys. Rev. A: At. Mol. Opt. Phys.},
  volume={88},
  pages={046103},
  year={2013},
  publisher={APS}
}

@article{morepast,
  title={Tracing the past of a quantum particle},
  author = {Vaidman, L.},
  journal={Phys. Rev. A},
  volume={89},
  pages={024102},
  year={2014},
  publisher={APS}
}

@article{Bart,
  title={One-state vector formalism for the evolution of a quantum state through nested {M}ach-{Z}ehnder interferometers},
  author = {Bartkiewicz, K. and {\v{C}}ernoch, A. and Jav{\u u}rek, D. and Lemr, K. and Soubusta, J. and Svozil{\'\i}k, J.},
  journal={Phys. Rev. A},
  volume={91},
  pages={012103},
  year={2015},
  publisher={APS}
}

@article{BartCom,
  title={Comment on “{O}ne-state vector formalism for the evolution of a quantum state through nested {M}ach-{Z}ehnder interferometers”},
  author = {Vaidman, L.},
  journal={Phys. Rev. A},
  volume={93},
  pages={036103},
  year={2016},
  publisher={APS}
}

@article{Poto,
  title={Which-way information in a nested {M}ach-{Z}ehnder interferometer},
  author = {Poto{\v{c}}ek, V. and Ferenczi, G.},
  journal={Phys. Rev. A},
  volume={92},
  pages={023829},
  year={2015},
  publisher={APS}
}

@article{PotoCom,
  title={Comment on “{W}hich-way information in a nested {M}ach-{Z}ehnder interferometer”},
  author = {Vaidman, L.},
  journal={Phys. Rev. A},
  volume={93},
  pages={017801},
  year={2016},
  publisher={APS}
}

@article{Grif,
  title={Particle path through a nested {M}ach-{Z}ehnder interferometer},
  author = {Griffiths, R. B.},
  journal={Phys. Rev. A},
  volume={94},
  pages={032115},
  year={2016},
  publisher={APS}
}

@article{GrifRep,
  title={Comment on “{P}article path through a nested {M}ach-{Z}ehnder interferometer”},
  author = {Vaidman, L.},
  journal={Phys. Rev. A},
  volume={95},
  pages={066101},
  year={2017},
  publisher={APS}
}

@article{Hash,
  title={Two-state vector formalism and quantum interference},
  author = {Hashmi, F. and Li, F. and Zhu, S. and Zubairy, M. S.},
  journal={J. Phys. A: Math. Theor.},
  volume={49},
  pages={345302},
  year={2016},
  publisher={IOP Publishing}
}

@article{HashCom,
  title={Comment on ‘{T}wo-state vector formalism and quantum interference’},
  author = {Vaidman, L.},
  journal={J. Phys. A: Math. Theor.},
  volume={51},
  pages={068002},
  year={2018},
  publisher={IOP Publishing}
}

@article{HashComRep,
  title={Reply to the comment on ‘{T}wo-state vector formalism and quantum interference’},
  author = {Hashmi, F. and Li, F. and Zhu, S. and Zubairy, M. S.},
  journal={J. Phys. A: Math. Theor.},
  volume={51},
  pages={068001},
  year={2018},
  publisher={IOP Publishing}
}

@article{Dupr,
  title={Null weak values and the past of a quantum particle},
  author = {Duprey, Q. and Matzkin, A.},
  journal={Phys. Rev. A},
  volume={95},
  pages={032110},
  year={2017},
  publisher={APS}
}

@article{DuprCom,
  title={Comment on “{N}ull weak values and the past of a quantum particle”},
  author = {Sokolovski, D.},
  journal={Phys. Rev. A},
  volume={97},
  pages={046102},
  year={2018},
  publisher={APS}
}

@article{DuprComRep,
  title={Reply to “{C}omment on ‘{N}ull weak values and the past of a quantum particle”’},
  author = {Duprey, Q. and Matzkin, A.},
  journal={Phys. Rev. A},
  volume={97},
  pages={046103},
  year={2018},
  publisher={APS}
}

@article{Eli,
  title={Past of a particle in an entangled state},
  author = {Paneru, D. and Cohen, E.},
  journal={Int. J. Quantum Inf.},
  volume={15},
  pages={1740019},
  year={2017},
  publisher={World Scientific}
}

@article{Disapp,
  title={The case of the disappearing (and re-appearing) particle},
  author = {Aharonov, Y. and Cohen, E. and Landau, A. and Elitzur, A. C.},
  journal={Sci. Rep.},
  volume={7},
  pages={531},
  year={2017},
  publisher={Nature Publishing Group UK London}
}

@article{Sokol2,
  title={Path probabilities for consecutive measurements, and certain “quantum paradoxes”},
  author = {Sokolovski, D.},
  journal={Ann. Phys.},
  volume={397},
  pages={474},
  year={2018},
  publisher={Elsevier}
}

@article{ACWE,
  title={The weak reality that makes quantum phenomena more natural: Novel insights and experiments},
  author = {Aharonov, Y. and Cohen, E. and Waegell, M. and Elitzur, A. C.},
  journal={Entropy},
  volume={20},
  pages={854},
  year={2018},
  publisher={MDPI}
}

@article{Saldan,
  title={Interpreting a nested {M}ach-{Z}ehnder interferometer with classical optics},
  author = {Saldanha, P. L.},
  journal={Phys. Rev. A},
  volume={89},
  pages={033825},
  year={2014},
  publisher={APS}
}

@article{Jordan,
  title={Can a {D}ove prism change the past of a single photon?},
  author = {Alonso, M. A. and Jordan, A. N.},
  journal={Quant. Stud.: Math. Found.},
  volume={2},
  pages={255},
  year={2015},
  publisher={Springer}
}

@article{JordanCom,
  title={When photons are lying about where they have been},
  author = {Vaidman, L. and Tsutsui, I.},
  journal={Entropy},
  volume={20},
  pages={538},
  year={2018},
  publisher={MDPI}
}

@article{Sali,
  title={Commentary:"{A}sking photons where they have been"-without telling them what to say},
  author = {Salih, H.},
  journal={Front. Phys.},
  volume={3},
  pages={47},
  year={2015},
  publisher={Frontiers Media SA}
}

@article{SaliCom,
  title={Response: Commentary:"{A}sking photons where they have been"-without telling them what to say},
  author = {Danan, A. and Farfurnik, D. and Bar-Ad, S. and Vaidman, L.},
  journal={Front. Phys.},
  volume={3},
  pages={48},
  year={2015},
  publisher={Frontiers Media SA}
}

@article{China,
  title={An ideal experiment to determine the ‘past of a particle’ in the nested {M}ach-{Z}ehnder interferometer},
  author = {Li, F. and Hashmi, F. and Zhang, J. and Zhu, S.},
  journal={Chin. Phys. Lett.},
  volume={32},
  pages={050303},
  year={2015},
  publisher={IOP Publishing}
}

@article{ChinaCom,
  title={An improved experiment to determine the ‘past of a particle’in the nested {M}ach--{Z}ehnder interferometer},
  author = {Ben-Israel, A. and Knips, L. and Dziewior, J. and Meinecke, J. and Danan, A. and Weinfurter, H. and Vaidman, L.},
  journal={Chin. Phys. Lett.},
  volume={34},
  pages={020301},
  year={2017},
  publisher={IOP Publishing}
}

@article{Nik,
  title={Paradox of photons disconnected trajectories being located by means of “weak measurements” in the nested {M}ach-{Z}ehnder interferometer},
  author = {Nikolaev, G. N.},
  journal={JETP Lett.},
  volume={105},
  pages={152},
  year={2017},
  publisher={Springer}
}

@article{NikRep,
  title={A Comment on "{P}aradox of photons disconnected trajectories being located by means of 'weak measurements' in the nested {M}ach-{Z}ehnder interferometer"({JETP L}etters 105, 152 (2017))},
  author = {Vaidman, L.},
  journal={JETP Lett.},
  volume={105},
  pages={473},
  year={2017}
}

@article{NikRR,
  title={Response to the {C}omment on" {P}aradox of Photons Disconnected Trajectories Being Located by Means of "Weak Measurements" in the Nested {M}ach-{Z}ehnder Interferometer"({JETP L}etters 105, 152 (2017))},
  author = {Nikolaev, G. N.},
  journal={JETP Lett.},
  volume={105},
  pages={475},
  year={2017}
}

@article{Sok,
  title={Asking photons where they have been in plain language},
  author = {Sokolovski, D.},
  journal={Phys. Lett. A},
  volume={381},
  pages={227},
  year={2017},
  publisher={Elsevier}
}

@article{SokCom,
  title={A Comment on "{A}sking photons where they have been in plain language"},
  author = {Vaidman, L.},
  journal={arXiv preprint arXiv:1703.03615},
  year={2017}
}

@article{SokComRep,
  title={Reply to {L}. {V}aidman's comment on "{A}sking photons where they have been in plain language"},
  author = {Sokolovski, D.},
  journal={arXiv preprint arXiv:1704.02172},
  year={2017}
}

@article{Berge,
  title={Past of a quantum particle revisited},
  author = {Englert, B. and Horia, K. and Dai, J. and Len, Y. L. and Ng, H. K.},
  journal={Phys. Rev. A},
  volume={96},
  pages={022126},
  year={2017},
  publisher={APS}
}

@article{BergeCom,
  title={Comment on “{P}ast of a quantum particle revisited”},
  author = {Peleg, U. and Vaidman, L.},
  journal={Phys. Rev. A},
  volume={99},
  pages={026103},
  year={2019},
  publisher={APS}
}

@article{BergeRep,
  title={Reply to “{C}omment on ‘{P}ast of a quantum particle revisited’”},
  author = {Englert, B. and Horia, K. and Dai, J. and Len, Y. L. and Ng, H. K.},
  journal={Phys. Rev. A},
  volume={99},
  pages={026104},
  year={2019},
  publisher={APS}
}

@article{Wiesn,
  title={Spectra in nested {M}ach--{Z}ehnder interferometer experiments},
  author = {Wie{\'s}niak, M.},
  journal={Phys. Lett. A},
  volume={382},
  pages={2565},
  year={2018},
  publisher={Elsevier}
}

@article{Yuan1,
  title={Three-path interference of a photon and reexamination of the nested {M}ach-{Z}ehnder interferometer},
  author = {Yuan, Q. and Feng, X.},
  journal={Phys. Rev. A},
  volume={99},
  pages={053805},
  year={2019},
  publisher={APS}
}

@article{Lemmel,
  title={Quantifying the presence of a neutron in the paths of an interferometer},
  author = {Lemmel, H. and Geerits, N. and Danner, A. and Hofmann, H. F. and Sponar, S.},
  journal={Phys. Rev. Res.},
  volume={4},
  pages={023075},
  year={2022},
  publisher={APS}
}

@article{Hance,
  title={Weak values and the past of a quantum particle},
  author = {Hance, J. R. and Rarity, J. and Ladyman, J.},
  journal={Phys. Rev. Res.},
  volume={5},
  pages={023048},
  year={2023},
  publisher={APS}
}

@article{HanceCom,
  title={Comment on “{W}eak values and the past of a quantum particle”},
  author = {Vaidman, L.},
  journal={Phys. Rev. Res.},
  volume={5},
  pages={048001},
  year={2023},
  publisher={APS}
}

@article{Danner,
  title={Simultaneous path weak-measurements in neutron interferometry},
  author = {Danner, A. and Masiello, I. V. and Dvorak, A. and Kersten, W. and Lemmel, H. and Wagner, R. and Hasegawa, Y.},
  journal={Sci. Rep.},
  volume={14},
  pages={25994},
  year={2024},
  publisher={Nature Publishing Group UK London}
}

@article{Kim,
  title={Which-path information of path-polarization hybrid state in polarization-based nested {M}ach-{Z}ehnder interferometer},
  author = {Kim, D. and Kim, M. and Moon, H. S.},
  journal={Sci. Rep.},
  volume={15},
  pages={12664},
  year={2025},
  publisher={Nature Publishing Group UK London}
}

@article{vaidman2024,
  title={Lying particles},
  author = {Vaidman, L.},
  journal={Front. Quantum Sci. Technol.},
  volume={3},
  pages={1362235},
  year={2024},
  publisher={Frontiers Media SA}
}

@article{Bhati,
  title={Do weak values capture the complete truth about the past of a quantum particle?},
  author = {Bhati, R. S. and Arvind},
  journal={Phys. Lett. A},
  volume={429},
  pages={127955},
  year={2022},
  publisher={Elsevier}
}

@article{PhotonsLyingAgain,
  title={Photons are lying about where they have been, again},
  author = {Reznik, G. and Versmold, C. and Dziewior, J. and Huber, F. and Bagchi, S. and Weinfurter, H. and Dressel, J. and Vaidman, L.},
  journal={Phys. Lett. A},
  volume={470},
  pages={128782},
  year={2023},
  publisher={Elsevier}
}

@article{Uranga,
  title={Quantum Weak Values and the “Which Way?” Question},
  author = {Uranga, A. and Akhmatskaya, E. and Sokolovski, D.},
  journal={Entropy},
  volume={27},
  pages={259},
  year={2025},
  publisher={MDPI}
}

@article{Dajka,
  title={Faint trace of a particle in a noisy {V}aidman three-path interferometer},
  author = {Dajka, J.},
  journal={Sci. Rep.},
  volume={11},
  pages={1123},
  year={2021},
  publisher={Nature Publishing Group UK London}
}

@article{Terris,
  title={Weak particle presence},
  author = {Terris, B.},
  journal={Found. Phys.},
  volume={55},
  pages={26},
  year={2025},
  publisher={Springer}
}

@article{Bernardo,
  title={How a single particle simultaneously modifies the physical reality of two distant others: a quantum nonlocality and weak value study},
  author = {de Lima Bernardo, B. and Canabarro, A. and Azevedo, S.},
  journal={Sci. Rep.},
  volume={7},
  pages={39767},
  year={2017},
  publisher={Nature Publishing Group UK London}
}

@article{YuanPhotons,
  title={Photons can tell “contradictory” answer about where they have been},
  author = {Yuan, Q. and Feng, X.},
  journal={Eur. Phys. J. Plus},
  volume={138},
  pages={70},
  year={2023},
  publisher={Springer}
}

@article{YuanPhotonsCom,
  title={Comment on “{P}hotons can tell ‘contradictory’ answer about where they have been”},
  author = {Reznik, G. and Versmold, C. and Dziewior, J. and Huber, F. and Weinfurter, H. and Dressel, J. and Vaidman, L.},
  journal={Eur. Phys. J. Plus},
  volume={139},
  pages={1},
  year={2024},
  publisher={Springer}
}

@incollection{McQueen,
  title={How the many worlds interpretation brings common sense to paradoxical quantum experiments},
  author = {McQueen, K. J. and Vaidman, L.},
  booktitle={Scientific challenges to common sense philosophy},
  pages={40--60},
  year={2020},
  publisher={Routledge}
}

@inproceedings{Geppert,
  title={Asking neutrons where they have been},
  author = {Geppert, H. and Denkmayr, T. and Lemmel, H. and Hasegawa, Y. and others},
  booktitle={J. Phys. Conf. Ser.},
  volume={1316},
  pages={012002},
  year={2019},
  organization={IOP Publishing}
}

@article{Waegell2023,
  title={Quantum reality with negative-mass particles},
  author = {Waegell, M. and Cohen, E. and Elitzur, A. and Tollaksen, J. and Aharonov, Y.},
  journal={Proc. Natl. Acad. Sci. U.S.A.},
  volume={120},
  pages={e2018437120},
  year={2023},
  publisher={National Academy of Sciences}
}

@article{Saeed,
  title={Past of a Quantum particle: An atom interferometric based study},
  author = {Saeed, M. H. and Awan, H. and Imran, M. and Ikram, M. and others},
  journal={Opt. Commun.},
  volume={520},
  pages={128536},
  year={2022},
  publisher={Elsevier}
}

@article{YuanHiding,
  title={Hiding the which-path information of photons},
  author = {Guo, J. and Yuan, Q. and Wu, Y. and Zhang, W.},
  journal={Phys. Rev. A},
  volume={106},
  pages={022210},
  year={2022},
  publisher={APS}
}

@article{ABL,
  title={Time symmetry in the quantum process of measurement},
  author = {Aharonov, Y. and Bergmann, P. G. and Lebowitz, J. L.},
  journal={Phys. Rev.},
  volume={134},
  pages={B1410},
  year={1964},
  publisher={APS}
}

@article{salcom,
  title = {Comment on ``{A}ssociation between quantum paradoxes based on weak values and a realistic interpretation of quantum measurements''},
  author = {Seoane, J. J. and Oianguren-Asua, X. and Sol\'e, A. and Oriols, X.},
  journal = {Phys. Rev. A},
  volume = {111},
  issue = {6},
  pages = {066201},
  numpages = {5},
  year = {2025},
  publisher = {American Physical Society},
}

@article{salrep,
  title = {Reply to ``{C}omment on `{A}ssociation between quantum paradoxes based on weak values and a realistic interpretation of quantum measurements' ''},
  author = {Aredes, A. M. and Saldanha, P. L.},
  journal = {Phys. Rev. A},
  volume = {111},
  issue = {6},
  pages = {066202},
  year = {2025},
  publisher = {American Physical Society},
}

@article{prodFailureExperimental,
  title={Experimental extraction of nonlocal weak values for demonstrating the failure of a product rule},
  author={Xu, Xiao-Ye and Pan, Wei-Wei and Kedem, Yaron and Wang, Qin-Qin and Sun, Kai and Xu, Jin-Shi and Han, Yong-Jian and Chen, Geng and Li, Chuan-Feng and Guo, Guang-Can},
  journal={Opt. Lett.},
  volume={45},
  number={7},
  pages={1715},
  year={2020},
  publisher={Optical Society of America}
}

@article{prl2026,
  title = {Interferometric Amplification and Suppression of External Beam Shifts},
  author = {Versmold, Carlotta and Dziewior, Jan and Huber, Florian and K\"oster, Elina and Reznik, Gregory and Vaidman, Lev and Weinfurter, Harald},
  journal = {Phys. Rev. Lett.},
  volume = {135},
  pages = {253802},
  numpages = {8},
  year = {2025},
  publisher = {American Physical Society},
}

\end{document}